%% file: conference_101719.tex
\documentclass[conference]{IEEEtran}
\IEEEoverridecommandlockouts
\usepackage{cite}
\usepackage{amsmath,amssymb,amsfonts}
\usepackage{algorithmic}
\usepackage{graphicx}
\usepackage{textcomp}
\usepackage{xcolor}

\usepackage{colortbl}
\usepackage{tcolorbox}
\usepackage{listings}
\usepackage{makecell}
\usepackage{multirow}
\usepackage{hyperref} 
\usepackage{fbox}
\usepackage{minibox}
\usepackage{balance}
\usepackage{enumitem}
\usepackage{xspace}
\usepackage{booktabs}
\usepackage{threeparttable}
\usepackage[]{collab}
\usepackage{url}

\def\BibTeX{{\rm B\kern-.05em{\sc i\kern-.025em b}\kern-.08em
    T\kern-.1667em\lower.7ex\hbox{E}\kern-.125emX}}

\newcommand\cloud{CloudA\xspace}
\newcommand\nm{LoFI\xspace}

\newcommand\datasetA{FIBench\xspace}
\newcommand\datasetB{Industry\xspace}
\newcommand\param{``$*$''\xspace}

\newcommand{\etc}{{\em etc}\xspace}
\newcommand{\ie}{{\em i.e.},\xspace}
\newcommand{\eg}{{\em e.g.},\xspace}

\newcommand{\blue}[1]{\textcolor{blue}{#1}}

\definecolor{ballblue}{rgb}{0.13, 0.67, 0.8}
\definecolor{jcpink}{RGB}{255, 0, 96}
\collabAuthor{jy}{ballblue}{Jinyang Liu}
\collabAuthor{jc}{jcpink}{JC}
\collabAuthor{jz}{magenta}{Jiazhen Gu}
\collabAuthor{zh}{green}{Zhihan Jiang}
\collabAuthor{jj}{purple}{Junjie}
\collabAuthor{zb}{teal}{Zhuangbin}
\collabAuthor{yc}{blue}{Yichen}

\begin{document}

\title{Demystifying and Extracting Fault-indicating Information from Logs for Failure Diagnosis}

\author{\large Junjie Huang\IEEEauthorrefmark{1}, Zhihan Jiang\IEEEauthorrefmark{1}, Jinyang Liu\IEEEauthorrefmark{1}, Yintong Huo\IEEEauthorrefmark{1}, Jiazhen Gu\IEEEauthorrefmark{1}, Zhuangbin Chen\IEEEauthorrefmark{2}\thanks{\hspace{-2ex}\IEEEauthorrefmark{2}Zhuangbin Chen is the corresponding author.},\\ Cong Feng\IEEEauthorrefmark{3}, Hui Dong\IEEEauthorrefmark{3}, Zengyin Yang\IEEEauthorrefmark{3}, Michael R. Lyu\IEEEauthorrefmark{1}\\
\IEEEauthorblockA{
\IEEEauthorrefmark{1}The Chinese University of Hong Kong, China, \{jjhuang23, zhjiang22, jyliu, ythuo, jiazhengu, lyu\}@cse.cuhk.edu.hk\\
\IEEEauthorrefmark{2}Sun Yat-sen University, China, \{chenzhb36@mail.sysu.edu.cn\}\\
\IEEEauthorrefmark{3}Huawei Cloud Computing Technology Co., Ltd, China, \{fengcong5, donghui25, yangzengyin\}@huawei.com\\
}

}

\maketitle
\begin{abstract}

Logs are imperative in the maintenance of online service systems, which often encompass important information for effective failure mitigation.
While existing anomaly detection methodologies facilitate the identification of anomalous logs within extensive runtime data, manual investigation of log messages by engineers remains essential to comprehend faults, which is labor-intensive and error-prone.
Upon examining the log-based troubleshooting practices at \cloud\footnote{Due to the company policy, we anonymize the name as \cloud.}, we find that engineers typically prioritize two categories of log information for diagnosis.
These include \textit{fault-indicating descriptions}, which record abnormal system events, and \textit{fault-indicating parameters}, which specify the associated entities.
Motivated by this finding, we propose an approach to automatically extract such fault-indicating information from logs for fault diagnosis, named \nm.
\nm comprises two key stages.
In the first stage, \nm performs coarse-grained filtering to collect logs related to the faults based on semantic similarity. 
In the second stage, \nm leverages a pre-trained language model with a novel prompt-based tuning method to extract fine-grained information of interest from the collected logs.
We evaluate \nm on logs collected from Apache Spark and an industrial dataset from \cloud.
The experimental results demonstrate that \nm outperforms all baseline methods by a significant margin, achieving an absolute improvement of 25.8\textasciitilde 37.9 in F1 over the best baseline method, ChatGPT.
This highlights the effectiveness of \nm in recognizing fault-indicating information.
Furthermore, the successful deployment of \nm at \cloud and user studies validate the utility of our method\footnote{The code and data are available at \url{https://github.com/Jun-jie-Huang/LoFI}}.

\end{abstract}

\begin{IEEEkeywords}
log analysis, failure diagnosis, fault-indicating information, cloud service system
\end{IEEEkeywords}

\input{content/01_introduction}

\input{content/02_background}

\input{content/03_empirical_study}

\input{content/04_method}

\input{content/05_experiment}

\input{content/06_discussion}

\input{content/07_related_work}

\input{content/08_conclusion}

\balance

\bibliographystyle{IEEEtran}
\bibliography{reference}

\end{document}

%% file: content/01_introduction.tex
\vspace{-0.4em}
\section{Introduction}\label{sec:intro}
\vspace{-0.2em}

As software systems grow in complexity, especially for online services with hundreds of distributed components serving global users, the challenge of preventing failures intensifies.
Despite extensive efforts, these systems still encounter large-scale unplanned interruptions and service quality degradation~\cite{aws, google, azure,huang2024faultprofit}.
To enhance user experience and minimize economic losses, IT companies must promptly and effectively respond to failures, thereby ensuring reliability of their software.

Logs, which document a variety of software runtime events, have been widely acknowledged as a crucial resource for diagnosing failures in online service systems~\cite{li2023exploring,li2024go,zhou2019latent, zhang2021onion}.
For example, an empirical study on a commercial bank service revealed that at least 31\% of failure diagnosis practices depend on logs~\cite{zhao2021empiricalLogAD}. 
However, the sheer scale and complexity of modern software lead to the generation of massive logs from multiple services, making it challenging for engineers to promptly examine logs and gain actionable insights about faults~\cite{lin2016logclustering,chen2024tracemesh}.

To facilitate rapid diagnosis, various log analysis approaches have been proposed. These approaches strive to identify and extract valuable information from massive log data in diverse granularities, aiming to reduce manual efforts in fault localization and investigation~\cite{he2021survey, zhao2021empiricalLogAD, he2022empirical,jiang2024lilac}.
Although they have made significant progress, the information they yield can still be extensive or irrelevant to faults, hindering the direct usage for system troubleshooting.
For example, log-based anomaly detection~\cite{he2016experience,chen2021experience,he2021survey,liu2023scalable} can identify anomalous log sessions (\eg an input list of logs that the model deems as anomalous) from a large volume of streaming logs. 
However, even though the identified sessions are small in time (\eg 10s), they could still contain hundreds of logs, with only a small portion being relevant to the fault~\cite{huo2023evlog}.
Log clustering~\cite{lin2016logclustering,zhao2020understanding,zhang2021onion} takes one step further to filter out irrelevant logs and localize the incident-indicating ones.
This is done by first grouping similar logs and then selecting the representatives.
However, the semi-structured nature of log messages can make them intricate and include redundant information about system execution.
As a result, it requires additional manual efforts to comprehend various parts of a log message.
Log parsing~\cite{he2017drain,zhu2019toolslogparser,jiang2024large}, on the other hand, can extract different elements of a raw log, such as log events and parameters. 
Yet, it remains unknown whether an event describes a critical failure symptom or a parameter captures the faulty component (e.g., device or VM). 
Semantic-aware log parsers~\cite{li2023valb,jiang2023llmparser,huang2024ulog} take a step further to identify parameter types (\eg object ID and type indicator) during parsing. However, the identified parameter types cannot always relate to faulty components.
As a result, there is still a lack of tools that can automatically extract more precise and crucial information from logs to guide engineers in taking immediate actions for fault diagnosis.

\begin{figure*}[t]
    \centering
    \includegraphics[width=1.8\columnwidth]{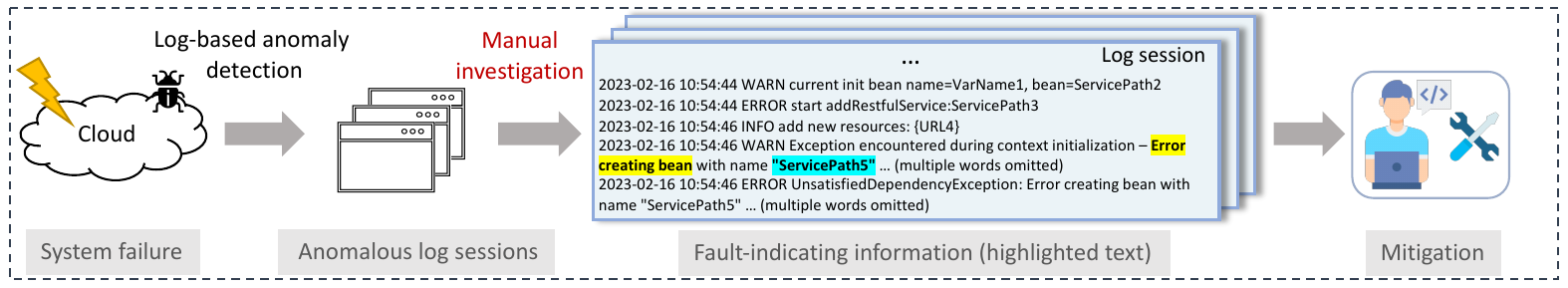}
    \vspace{-1em}
    \caption{The workflow of failure diagnosis with logs.}
    \vspace{-1.6em}
    \label{fig: workflow}
\end{figure*}

In this paper, we propose to extract \emph{fault-indicating information} from logs, namely the parts of logs that convey direct and valuable insights into system faults.
Utilizing this information, engineers can quickly understand the underlying issues and examine the faulty components for effective fault mitigation. 
To figure out what information is fault-indicating, we first conduct a preliminary study based on the fault mitigation practices at \cloud, a top-tier global cloud vendor.
By analyzing historical fault diagnosis reports and the associated logs, we summarize two categories of fault-indicating information that are frequently employed in troubleshooting, \ie fault-indicating descriptions (FID) and fault-indicating parameters (FIP).
FID describes the symptoms of a fault, which is further divided into four subtypes including error message, missing component, abnormal behavior, and wrong status.
While FIP pinpoints the exact location of the fault that demands investigation, which have three typical subtypes, \ie address, component ID, and parameter name.
As illustrated by the example in Figure~\ref{fig: workflow}, the FID of the anomalous session is ``Error creating bean'', and the FIP is ``ServicePath5''.

Designing an automated tool to extract fault-indicating information from log messages presents two key challenges. 
(1) The first challenge relates to the vast volume of log messages generated by modern online service systems. 
For example, these systems can produce up to 200 million lines of logs per hour~\cite{wang2022spine}.
Although anomaly detection offers potential benefits, the number of logs requiring investigation in an anomalous session can still be substantial, often ranging from hundreds to thousands~\cite{zhu2019toolslogparser}.
Our study reveals that, for log messages in an anomalous session, only a small percentage (\ie 1.7\%) contains information indicative of faults~\cite{chuah2010diagnosing, zhang2021onion}.
(2) The second challenge involves dealing with noisy semantics in log messages, as engineers often write detailed logging statements to provide supplementary runtime information~\cite{yuan2012logenhancer}.
Consequently, the content of log messages tends to be verbose, with only a small portion of keywords or phrases describing the underlying issues.
Our study reveals that only approximately 14.1\% of the words in log messages indicate a fault or issue, posing significant challenges on the identification of target information.
Overcoming these two challenges is crucial in developing an efficient and accurate tool for extracting fault-indicating information from log messages.

To address the challenges, we propose \nm, a two-stage approach for efficient and accurate extraction of \textbf{Lo}g \textbf{F}ault-\textbf{I}ndicating information, \ie FID and FIP, to aid in fault diagnosis and reduce manual mitigation effort.
\nm mainly consists of two stages: \textit{log selection} and \textit{prompt-based extraction}.
In the first stage, \nm uses a coarse-grained filtering mechanism to select logs potentially relevant to faults, thereby reducing noise.
This is done by first collecting logs with severe logging levels, and then incorporating more relevant logs based on their semantic similarity.
By eliminating less important logs in anomalous sessions, the extraction of fine-grained information in the next stage becomes more effective and efficient.
In the second stage, \nm extracts fault-indicating information from the selected log messages by tuning a pre-trained language model (PLM).
Inspired by the success of prompt learning~\cite{le2023logppt}, we adopt a question-answering schema to query the PLM, instructing it to iteratively yield FID and FIP. 
This enables us to design specific prompts that convey precise instructions about the desired information.
As a result, the PLM can ignore noises within log messages and provide more accurate results.

We evaluate \nm on logs from Apache Spark, a distributed data processing system~\cite{zaharia2016apache}, and an industrial dataset from \cloud.
Given that the extraction of fault-indicating information is a novel task in log analysis, we created a benchmark dataset based on Apache Spark, namely \datasetA.
This is achieved by initially injecting faults~\cite{lee2023heterogeneous} in to the system and then manually pinpointing fault-indicating information from the injection history.
To further validate \nm's practical significance, we also conduct experiments on an industrial dataset from \cloud, named \datasetB.
The dataset is collected from historical postmortem reports and validated by experienced on-site engineers.
Overall, \nm achieves an F1 score of 87.4/80.6 for FID/FIP extraction on \datasetA and 72.2/62.8 on \datasetB, significantly outperforming all baseline methods (81\% on average higher than the best baseline method, ChatGPT-ICL~\cite{chatgpt}).
Our ablation experiments and user study further confirm the effectiveness and practical usefulness of our method.

This work makes the following main contributions:
\begin{itemize}[leftmargin=*, topsep=0pt]
    \item We conduct a preliminary study based on six-month-long fault diagnosis reports at \cloud (\S\ref{sec:empiral-study}), and summarise two categories of fault-indicating information that provides valuable insights to on-site engineers, \ie fault-indicating description (FID) and fault-indicating parameter (FIP).
    
    \item We propose \nm, an approach to automatically extract fault-indicating information from anomalous log sessions (\S\ref{sec:method}). \nm utilizes a novel prompt-based tuning method to effectively learn semantic information from logs with limited training data. 
    
    \item Extensive experiments are conducted on datasets collected from Apache Spark and an industrial system (\S\ref{sec:evaluation}). The results show that \nm outperforms state-of-the-art methods, including ChatGPT, by a large margin. We also demonstrate the practical value of \nm through a user study at \cloud.
\end{itemize}

%% file: content/02_background.tex
\vspace{-0.4em}
\section{Background and Motivation}
\vspace{-0.2em}
\subsection{Log Analysis for Fault Diagnosis}
\vspace{-0.2em}

Online service systems typically generate logs to record runtime status for troubleshooting. A log entry generally comprises three parts: a \emph{timestamp} recording the time that the entry is recorded, a \emph{level} indicating the severity of a log entry (\eg INFO, WARN and ERROR), and \emph{content} containing human-readable information that describes the specific system status. 
Logs serve as an important data source for failure detection and diagnosis in software systems~\cite{lin2016logclustering}. 

However, manually investigating a huge number of logs is labor-intensive and time-consuming, which can lead to prolonged fault mitigation time. Thus many methods have been proposed to identify and understand useful information from logs to expedite the process. \textbf{Log parsing} is the first step to enable automated log analysis~\cite{zhu2019toolslogparser}, which converts raw logs into \emph{log templates} describing events and \emph{variables} recording dynamic runtime information. Since the log parsing results cannot be directly used for diagnosis~\cite{li2023valb},
\textbf{log-based anomaly detection} methods~\cite{xu2009largescale, lin2016logclustering} are proposed to identify abnormal system behaviors from logs and reduce millions of logs to a small window for engineers to investigate. However, as revealed by recent studies~\cite{dogga2019debuggingAssistant, meng2023logsummary}, on-site engineers still have to manually investigate the runtime failure by reading tens or even hundreds of raw and noisy logs before taking further troubleshooting measures. 
Considering the defects of existing methods, there is an urgent need to automatically identify the detailed and crucial information from logs in order to assist in-time diagnosis and rapid mitigation.

\vspace{-0.5em}
\subsection{A Motivating Example}

In this section, we introduce the current practices of log-based fault diagnosis at \cloud, which motivate this work.
Figure \ref{fig: workflow} shows a typical workflow to diagnose a fault at \cloud. 
The fault in this case was caused by a flawed configuration update to the job scheduling module of service X, leading to degraded performance and a decrease in successful requests. 
Upon detecting the failure, on-site engineers followed a systematic approach. They first identified relevant services based on their experience, including the manage-board service, monitoring service, and database. 
They then retrieved logs from these services for diagnosis. In this process, anomaly detection methodologies can help pinpoint anomalous log sessions.
Next, they conducted a coarse-grained search to locate critical log messages containing essential information about the fault, using keywords  ``kill'', ``fail'', ``error'', and ``exception''. 
Subsequently, a fine-grained analysis was performed by carefully reading the logs to understand the issue and determine the root cause. 
For example, by reading logs from the manage-board service, engineers discovered that the bean configuration creation failed, specifically in the problematic instance ServicePath5. 
During investigation, this process was repeated for all services to ensure a comprehensive analysis. 
Finally, they identified misconfiguration as the root cause and proceeded to fix the fault and restart the service.

Based on the aforementioned process, we can observe that while log-based anomaly detection expedites the coarse-grained search to identify anomalous log sessions, a more detailed and precise fine-grained investigation remains necessary for fault diagnosis.
The growing scale and complexity of online service systems mean that a single fault may encompass numerous log sessions, each containing tens or even thousands of log messages. 
Consequently, addressing these issues requires significant manual effort.
Based on the on-site engineers of \cloud, manual fault investigation through logs accounts for nearly two-thirds of the overall fault handling time.
This observation aligns with the findings from recent studies~\cite{zhang2021onion, li2021fogofwar}.
Thus, this work aims to address the challenge of automatically extracting fine-grained information from log sessions to enhance the process of fault diagnosis.

%% file: content/03_empirical_study.tex
\input{tables/log_example}

\section{Fault-indicating Information in Logs}\label{sec:empiral-study}

In this section, we aim to summarize the essential information capable of assisting engineers in taking appropriate actions and pinpointing the fault, \ie \textit{fault-indicating information}.
To this end, we conduct a preliminary study to examine how engineers leverage logs in their fault handling processes.
The study comprises three steps: \textit{dataset preparation}, \textit{manual investigation}, and \textit{result analysis}.

\noindent\textbf{Dataset Preparation.}
We start by collecting historical faults and the corresponding diagnostic reports from service X, a large-scale online service system of \cloud. 
Service X adopts the microservice architecture with rich functionalities such as user management, analytics, resource scheduling, logging and monitoring, \etc.
The data span from 2022-09-03 to 2023-03-02, resulting in a total of 88 faults.
These faults cover diverse root causes, including network disconnection, device failures, configuration errors, \etc. 
For each fault, on-site engineers documented diagnosis details such as affected components, associated log files, and a fault summary that included diagnosis process, root causes, and mitigation steps. 
We manually investigate these reports to gain insights into how log messages are utilized during fault mitigation.

\noindent\textbf{Manual Investigation.}
The goal of the manual investigation is to identify fault-indicating information that aids in fault diagnosis. 
Generally, we pinpoint such information by manually analysing diagnosis reports and corresponding log sessions to identify the log segments that are notified in the reports. 
The motivation of the pinpointing strategy is that only important details will be recorded in the reports, where the notified events or components are more likely to indicate faults.

Specifically, we first collect logs that are generated from around ten minutes before and after each fault given the recorded timestamps.
These logs are likely to cover all system events associated with the fault.
We then examine these logs and the diagnostic report to identify the specific log messages that are directly related to the fault or explicitly referred to in the mitigation step.
For example, if a mitigation step mentions restarting a virtual machine with ID=6afd89eh, we will mark this ID. 
This process allows us to locate relevant log information used by on-site engineers in fault resolution.
Two authors conduct the manual investigation separately for all collected faults. Subsequently, two senior on-site engineers from \cloud review the results for correctness. Any discrepancies are resolved through discussion until a consensus is reached.

\noindent\textbf{Result Analysis.}
From our manual investigation, we identified two main categories of fault-indicating information frequently used in the mitigation process: fault-indicating description (\textit{FID}), which primarily describes the fault, and fault-indicating parameter (\textit{FIP}), which denotes positions or components requiring further investigation.
Notably, fault-indicating parameters could only be extracted from 68 out of 88 faults, as the remaining faults do not explicitly mention parameters in the logs during mitigation. 
Within these two categories, we have further classified FID and FIP into four and three subtypes, respectively, based on their specific content as shown in Table~\ref{tab:empirial-desc-param-types}.
We introduce these subtypes as follows:

\begin{itemize}[leftmargin=*, topsep=0pt]
\item \textit{Error Message} directly describes a failed action or an exception raised from a software stack.
\item \textit{Missing Component} means some components are unavailable such as devices, tasks and hosts.
\item \textit{Abnormal Behavior} indicates the degraded performance of an application \eg HTTP timeout, slow response time.
\item \textit{Wrong Status} means a specific response code is incorporated to explain the wrong event, \eg  status code, error flags.
\item \textit{Address} includes a concrete URL of HTTP requests, IP address or paths to a folder. 
\item \textit{Component ID} records the index for a system component \eg job ID, task ID, service ID. 
\item \textit{Parameter Name} shows the key and value for a parameter \eg data name, user name.
\end{itemize}

Based on the identified fault-indicating information, we further investigate the presence of noise in the logs. 
Specifically, we compute the ratio of log messages containing either FID or FIP in their raw form. 
Our analysis reveals that, on average, only 1.7\% of the log messages contain the target information. 
Additionally, the fault-indicating words account for only 14.1\% of these messages' content.
These findings indicate a significant amount of noise in the logs, potentially hindering on-site engineers' ability to identify precise information needed to address the underlying faults.

In summary, we have identified two common categories of fault-indicating information in fault diagnosis: fault-indicating description, reflecting fault symptoms, and fault-indicating parameters, indicating faulty components.
Therefore, our focus is on extracting these two types of information.

%% file: tables/log_example.tex
\begin{table*}[h]
  \centering
    \caption{Categories of fault-indicating information in logs.}
\vspace{-1em}
    \resizebox{0.85\textwidth}{!}{
    \begin{threeparttable}
    \begin{tabular}{clrr}
    \toprule
    \multicolumn{1}{l}{Category} & Subtype & \multicolumn{1}{l}{Example} & \multicolumn{1}{l}{Number} \\
    \midrule
    \multicolumn{1}{c}{\multirow{4}[2]{*}{\shortstack{Description\\(FID)}}} & Error Message & \multicolumn{1}{l}{$\dots$\textbf{url detection error}!agent \underline{taskId:f292c7e596d5435d9b9e9b9f47e1f872}, retCode is empty} & 
    \multicolumn{1}{l}{32/88} \\
    & Missing Component &  \multicolumn{1}{l}{$\dots$execute template error, reason is \textbf{Host name must not be empty}}     & \multicolumn{1}{l}{20/88} \\
    & Abnormal Behavior &  \multicolumn{1}{l}{$\dots$reader request line for \underline{192.168.132.245:8080(https)} failed, \textbf{read line timed-out}} &  \multicolumn{1}{l}{24/88} \\
    & Wrong Status &   \multicolumn{1}{l}{$\dots$\textbf{httpCode is 404. requestEntity's type is GET}. requestEntity's url is \underline{/users/orders/task}. please check!}    &  \multicolumn{1}{l}{12/88} \\
    \midrule
    
    \multirow{3}{*}{\shortstack{Parameter\\(FIP)}} & Address  &     \multicolumn{1}{l}{$\dots$\textbf{httpCode is 404. requestEntity's type is GET}. requestEntity's url is \underline{/users/orders/task}. please check!}  & \multicolumn{1}{l}{15/68} \\
    & Component ID & \multicolumn{1}{l}{$\dots$\textbf{query $\dots$ failed}. \underline{historyid=51890bae-57c6-47a3-b37d-62df9d2f3c87}}      & \multicolumn{1}{l}{28/68} \\
    & Parameter Name &   \multicolumn{1}{l}{$\dots$\textbf{cannot get topicInfo for consumer by topic}: \underline{alarm\_and\_event\_data}}    & \multicolumn{1}{l}{25/68} \\
    \bottomrule
    \end{tabular}
    \begin{tablenotes}
        \small
        \item[1] Fault-indicating descriptions (FID) is marked in \textbf{BOLD} and fault-indicating parameters (FIP) is in \underline{UNDERLINE}.
    \end{tablenotes}
    \end{threeparttable}
}
\vspace{-4mm}
  \label{tab:empirial-desc-param-types}%
\end{table*}

%% file: content/04_method.tex
\vspace{-0.6em}
\section{Methodology}~\label{sec:method}
In this section, we describe our method \nm to automatically extract fault-indicating information from logs, \ie \emph{FID} and \emph{FIP} defined in \S~\ref{sec:empiral-study}. There are two major challenges in fault-indicating information extraction: 1) the log sessions produced by anomaly detection methods may still contain numerous, noisy log messages; 2) since each log message contains complex and redundant contents, it is hard to identify the fault-indicating information.  

\vspace{-0.4em}
\subsection{Problem Formulation}\label{sec:method-problem-formulation}
\vspace{-0.4em}
We formulate the problem of identifying fault-indicating information in logs as follows.
The input is an anomaly session with $n$ log messages $L = [l_1, l_2, \dots, l_n]$. Each log, represented as $l_i = [w_1, \dots, w_{m_i}]$, comprises a sequence of $m_i$ tokens.
The output is the fault-indicating information denoted as a tuple $(d, p)$, where $d = [w_1, \dots, w_{N_d}]$ represents the FID of system status and $p = [w_1, \dots, w_{N_p}]$ denotes the FIP for localization. 
$N_d$ and $N_p$ are the token numbers of FID and FIP, respectively.
The task aims to extract fault-indicating information $(d, p)$ to enable engineers to understand the faults and locations so that they can 
take appropriate actions. 

\vspace{-0.4em}
\subsection{Overview}
\vspace{-0.4em}
To overcome the aforementioned challenges, we propose a novel method called \nm, which is based on a pre-trained language model (PLM) and is designed to extract fault-indicating information from anomalous log sessions produced by preceding anomaly detectors. The main idea behind \nm is to utilize a large PLM to comprehend the semantics of log sessions and then extract the desired fault-indicating information by designing a prompt that accurately captures the intentions of on-site engineers.
Specifically, the \nm method is composed of two main stages: \textit{log selection} and \textit{prompt-based extraction}. In the log selection stage, \nm selects candidate logs that are likely to be related to the fault in a coarse-grained manner. This stage helps to filter out less relevant logs, thus enabling a more effective and efficient extraction process.
In the prompt-based extraction stage, \nm extracts FID and FIP from candidate log messages. This is accomplished by querying a fine-tuned PLM, such as UniXcoder~\cite{guo2022unixcoder}, with a well-designed prompt that precisely points to the fault-indicating information within the log messages. 
In this way, we allow the PLM to accurately return the desired information of on-site engineers.

\begin{figure}[t]
    \centering
    \includegraphics[width=0.80\columnwidth]{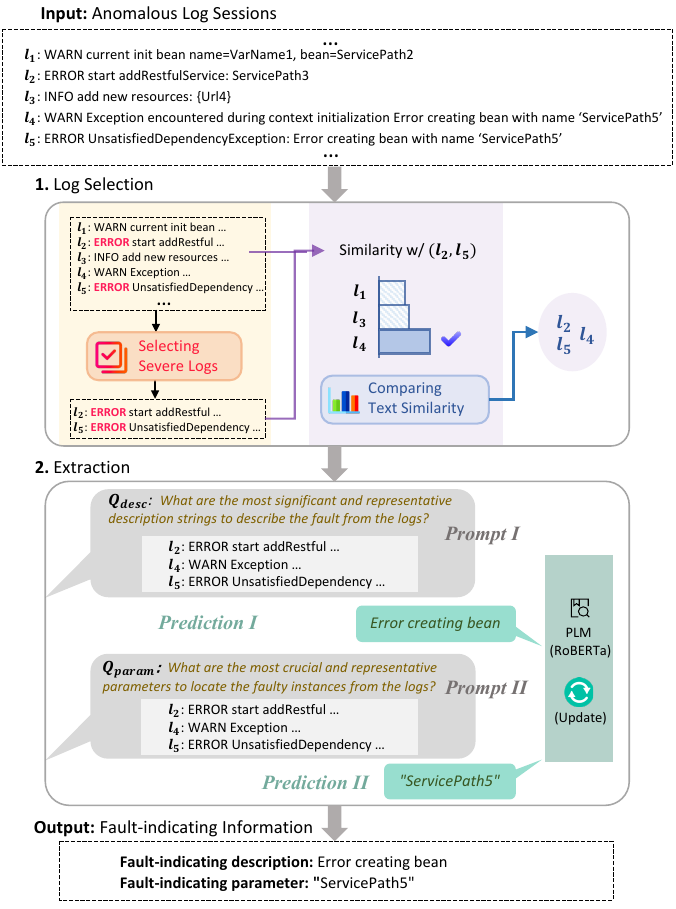}
    \vspace{-1em}
    \caption{The overall framework of \nm.}
    \label{fig:model-framework}
    \vspace{-2em}
\end{figure}

\vspace{-0.2em}
\subsection{Log Preprocessing}\label{sec:method-preprocess}
The preprocessing module converts raw log messages from a session into a formatted input for \nm. The input part of Figure~\ref{fig:model-framework} depicts an example of processed log messages. 
Given a raw log, the preprocessing module first uses regular expressions to split an unstructured log message into timestamp, logging level, and log content and then deduplicates log contents.
In our work, we keep the original log content for fault-indicating information extraction without parsing it into static log template and the corresponding dynamic variables~\cite{he2017drain}, since log parsing replaces common parameters with a unified identifier (\eg \param) which may mask informative variables (\eg device ids) and cut out the semantic relations between logs, leading to poor extraction performance.

\vspace{-0.4em}
\subsection{Log Selection}\label{sec:method-stage-1}
\vspace{-0.2em}
Sessions flagged as abnormal by upstream log-based anomaly detectors often contain a large number of logs, sometimes exceeding hundreds or thousands of lines~\cite{zhao2020understanding}. Understanding a large volume of logs is challenging for PLMs, as they often contain irrelevant information that complicates fault diagnosis. Additionally, PLMs have been shown to struggle with long texts~\cite{beltagy2020longformer}, emphasizing the need for a more concise input. 
To address the issues, we propose a log selection approach to filter out most irrelevant log messages to the fault, which includes two steps: \textit{level selection} (\S\ref{sec:method-stage-1-level}) and \textit{semantic selection} (\S\ref{sec:method-stage-1-semantic}).

\vspace{-0.5mm}
\subsubsection{Level Selection.}\label{sec:method-stage-1-level}
Our study in \S\ref{sec:empiral-study} shows that logs with more severe logging levels will be examined with a higher priority.
Therefore, we first apply \textit{level selection} to select logs with the highest logging level among those in a log session, which are denoted as~$L_{severe}$; and the left logs are denoted as $L_{mild}$.
The priority of logs is assigned by the standard ranking of logging levels defined in \textit{Log4j}~\cite{log4j}, \ie FATAL, ERROR, WARN, INFO, DEBUG, TRACE, Others. 

\vspace{-0.5mm}
\subsubsection{Semantic Selection.}\label{sec:method-stage-1-semantic}
However, relying solely on level selection can result in overlooking logs that, although not marked at a severe level, carry fault-indicating information. Those logs often have similar contents as severe logs, despite their less severe levels.
Thus, we apply \textit{semantic selection}, which involves calculating log embeddings and then performing similarity search to include these pertinent logs.

To produce log embeddings, we use UniXcoder~\cite{guo2022unixcoder}, a PLM trained on a mixture of natural language (NL) and code corpus. 
While alternative PLMs such as RoBERTa~\cite{liu2019roberta, le2023logppt} could also be considered, we believe UniXcoder is better suited to understand log messages that contain both NL and code-like parameters, enabling producing enhanced log representations.
Thus we follow~\cite{guo2022unixcoder} and use the default method to obtain log vectors by adopting the vector of the first token as the representation of the log message: $\mathbf{l_i} = \texttt{UniXcoder(}l_i\texttt{)}[0]$.

Upon obtaining log embeddings, we perform similarity search to include more relevant logs. 
For each log $l_i \in L_{mild}$, we compute its pairwise cosine similarity to the severe logs as $sim(l_i, l_j) = cosine(\mathbf{l_i}, \mathbf{l_j}), l_j \in L_{severe}$. We use cosine similarity as the distance metric since it measures the angle of two text vectors and is independent of their magnitude, making it suitable to measure textual semantic similarity~\cite{haering2021automatically}.
Then we take the maximum similarity as the final score $sim(l_i, L_{severe}) = \max_{l_i \in L_{severe}}(sim(l_i, l_j))$, as we aim to highlight the most similar logs in $L_{mild}$ to $L_{severe}$. Finally, we rank the logs by the score and take the top 10\% logs as $L_{similar}$. We merge $L_{severe}$ and $L_{similar}$ while preserving the original time order to create candidate logs. Despite the possibility of introducing noise, we still apply semantic selection since insufficient context could lead to further information loss. 
We show a comparison of selection methods in \S~\ref{sec:experiment-rq2}. 

\subsection{Prompt-based Extraction }\label{sec:method-stage-2} 
\vspace{-0.2em}
After log selection, we extract fault-indicating information from the candidate logs.
This is achieved by leveraging PLMs and prompt-tuning, which have been proven effective in recent log studies \cite{le2021NeuralLog, le2023logppt}. 
The essence behind this idea is twofold.
First, fault-indicating information has some similar semantic patterns, \eg using human-readable words to describe a failed operation or report a status code, which can be captured by PLMs through supervised signals. 
Second, the semantic patterns can vary across different logs, which is hard to detect by rule-based methods and cover all possible patterns.  
To implement this approach, we first choose a PLM (\S~\ref{sec:method-stage-2-plm}) and design prompt questions for FID and FIP (\S~\ref{sec:method-stage-2-prompt}). Then we combine the question and logs as input to the PLM to predict position spans of target fault-indicating information (\S~\ref{sec:method-stage-2-span-prediction}). 
Finally, we fine-tune the PLM with a few labeled examples to improve the performance on this task (\S~\ref{sec:method-stage-2-training}).

\subsubsection{Pre-trained Language Model (PLM)}\label{sec:method-stage-2-plm} PLMs~\cite{liu2019roberta} have shown remarkable abilities to understand the semantic meaning of logs~\cite{le2021NeuralLog, le2023logppt}. 
In this work, we employ UniXcoder~\cite{guo2022unixcoder} as our backbone for two main reasons. Firstly, UniXcoder is trained on a blend of NL and code corpus, making it more robust and capable of comprehending logs that contain similar parametric content.
Secondly, UniXcoder utilizes byte-level Byte-Pair Encoding (BPE)~\cite{sennrich2016bpe} for text tokenization, allowing it to avoid the out-of-vocabulary (OOV) problem by breaking OOV words into subwords.

\subsubsection{Prompt Tuning with Questions Answering}\label{sec:method-stage-2-prompt}
Prompt tuning is an effective method to apply PLM to downstream tasks by adding NL instructions to the input, which can better utilize the knowledge in PLMs~\cite{liu2023promptSurvey}. 
In this work, we transform fault-indicating information extraction into a question-answering task \cite{rajpurkar2016squad} by adding a prompt question. This approach enables UniXcoder to predict both FID and FIP from logs using a single unified model, eliminating the need for separate modules and enabling the sharing of common knowledge between the two types of information.

To effectively guide the PLMs, we design different prompt questions to identify FID and FIP. 
Specifically, for FID, the prompt question is \textit{``What are the most significant and representative description strings to describe the fault from the logs?''} For FIP, the prompt question is \textit{``What are the most crucial and representative parameters to locate the faulty instances from the logs?''}

\begin{figure}[t]
    \centering
    \includegraphics[width=0.92\columnwidth]{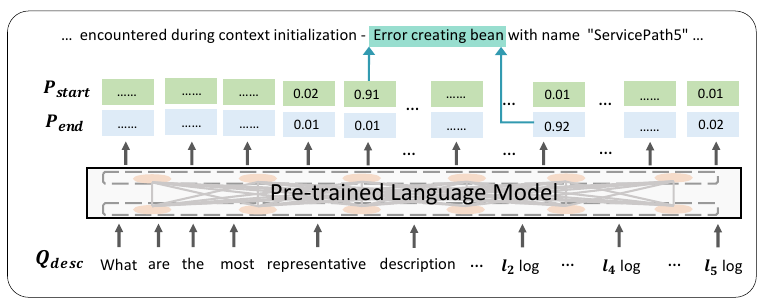}
    \vspace{-1em}
    \caption{Extracting fault-indicating information with the extraction module.}
    \label{fig:model-plm}
\vspace{-1.5em}
\end{figure}

\subsubsection{Predicting Span to Extract Fault-indicating Information}\label{sec:method-stage-2-span-prediction}
We use a span prediction mechanism~\cite{devlin2019bert} to extract FID and FIP spans with the PLM. This method, commonly used in question answering tasks~\cite{rajpurkar2016squad}, predicts the start and end positions of the target span.
By adding two additional vectors during fine-tuning, one for training a start position classifier and another for training an end position classifier, this method effectively aligns pre-training knowledge with our task requirements while maintaining a low parameter count. which is particularly beneficial when training data is scarce.

Figure \ref{fig:model-plm} shows the extraction process to obtain one span. Given a prompt question and logs $L_{severe}+L_{similar}$, we concatenate them with a special delimiter token~\texttt{[SEP]} to obtain a single packed input sequence $X$ to UniXcoder: 
\begin{equation}
\vspace{-0.2em}
\small
    X = \texttt{[CLS]}\; Question\; \texttt{[SEP]}\; l_1\; \texttt{[SEP]}\;  ...\;  l_i\; \texttt{[SEP]}
\vspace{-0.2em}
\end{equation}
The input sequence $X$ is then fed into UniXcoder to obtain hidden representations~$\mathbf{X} \in \mathbb{R}^{n \times d}$, where $n$ is the number of tokens and $d$ is the embedding dimension. We introduce a start vector~$\mathbf{s} \in \mathbb{R}^{d}$ and an end vector~$\mathbf{e} \in \mathbb{R}^{d}$ for prediction. The probability of token~$i$ being the start of the answer span is computed as a dot product between its representation $\mathbf{X}_{[i,:]}$ and $\mathbf{s}$ followed by a softmax function over all tokens: $P_{start} = \frac{e^{\mathbf{s} \cdot \mathbf{X}_{[i,:]} }}{\sum_j e^{\mathbf{s} \cdot \mathbf{X}_{[j,:]}}}$. Similarly, the probability of the end of a answer span can be computed by the analogous formula. The probability of a candidate span from position $i$ to $j$ is the product of start and end position probability: $P_{i, j} = P_i \cdot P_j$. Tokens in top-$k$ scoring spans where $j >= i$ are concatenated as the prediction.

\subsubsection{Training Objective}\label{sec:method-stage-2-training}
We use the start position $X_{start}$ and end position $X_{end}$ of ground truth FID or FIP as the target positions for each example $X$. The UniXcoder is trained to maximize the probability of the correct answer span, which is the sum of log-likelihoods of the correct start and end positions. The loss function is computed as follows:
\begin{equation}
\vspace{-0.2em}
\small
    Loss = - \frac{1}{N} \sum_N (\log P_{X_{start}} + \log P_{X_{end}}),
\vspace{-0.2em}
\end{equation}
where $N$ is the number of labeled training examples.

\vspace{-2mm}
\subsection{Online Extraction}\label{sec:method-online}
Identifying fault-indicating information from a large volume of logs can greatly assist engineers in monitoring system status and reducing diagnosis efforts in real-world software maintenance. Here, we apply \nm to extract FID/FIP in an online setting, where logs are continuously produced by software systems. To simulate the online phase, we first perform automatic log anomaly detection to prepare anomalous log sessions and then apply \nm to extract FID and FIP from the sessions. We take a simple yet effective method to detect anomalies with Decision Tree (DT)~\cite{wu2008dt}, which can efficiently scale to millions of logs in a short period of time \cite{he2016experience}. Specially, we parse the logs into log templates with Drain \cite{he2017drain} and compute a count feature vector for each session of logs. The count vector is subsequently used as input for DT, which is trained with 100 manually annotated anomalous sessions as positive examples. After anomaly detection, we use \nm to predict FID and FIP. Note that we only perform extraction for potentially anomalous log sessions, which is still efficient when scaled to a large volume of streaming logs.

%% file: content/05_experiment.tex
\vspace{-2mm}
\section{Evaluation}\label{sec:evaluation}
\vspace{-1mm}
We evaluate our method by answering the following research questions (RQs):
\begin{itemize}[leftmargin=*, topsep=0pt]
    \item RQ1: How effective is \nm in the \textit{offline} setting?
    \item RQ2: How log selection affects the results of \nm? 
    \item RQ3: How prompt-based tuning affects the results of \nm? 
    \item RQ4: How \nm helps SREs to diagnose in \textit{online} setting?
\end{itemize}

\vspace{-1.5mm}
\subsection{Experiment Designs}\label{sec:experiment-design}
\vspace{-1mm}
\subsubsection{Dataset}
To evaluate the effectiveness of \nm in extracting fault-indicating information from logs, we conduct experiments on two datasets, which are collected from a public software Apache Spark and a large-scale industrial system.

\noindent \textbf{\datasetA.}
We collect \datasetA from Apache Spark, a widely-used distributed framework for big data processing~\cite{zaharia2016apache}, which produces extensive log data to record its runtime information.
\cite{lee2023heterogeneous} conducted fault injection to Spark to gather logs in both normal and anomalous running states for log-based anomaly detection. They injected 21 types of faults, including network issues, process killing, \etc. However, the labeled anomalous log sections still contain numerous logs without FID and FIP. 
Building on the work~\cite{lee2023heterogeneous}, we further annotate the FID and FIP within these logs by examining the fault-injecting steps. Specifically, we focus on the logs containing entities that were first operated for fault injection. 
Next, by referring to the categories of fault-indicating information outlined in Table~\ref{tab:empirial-desc-param-types}, we can identify the FID and FIP in logs accurately.
For instance, if a process in a physical machine is terminated, we can associate the process ID and machine's IP address with the logs, thereby identifying the corresponding FID and FIP. Finally, we obtained 71 fault cases with explicit fault-indicating information.

\noindent \textbf{\datasetB.} To confirm \nm's practical significance, we also collect an industrial dataset with 88 fault cases from 33 microservices within the production system of \cloud. 
The detailed collection process of \datasetB can be found in \S\ref{sec:empiral-study}. Table \ref{tab:dataset-statistics} shows the statistics of our dataset.

\blue{
\begin{table}[htb]
\small
\centering
\vspace{-7.5mm}
\caption{Dataset statistics}
\vspace{-2.5mm}
\resizebox{0.40\textwidth}{!}{
\begin{tabular}{cccccc}
\toprule
Dataset & Total Logs & Logs per Session & Faults & FID & FIP  \\
\midrule
\textbf{\datasetA} & 1,225,287 & 39.9 & 71 & 71 & 37 \\
\textbf{\datasetB} & 2,721,013 & 64.3 & 88 & 88 & 68 \\
\bottomrule
\end{tabular}\label{tab:dataset-statistics}
}
\vspace{-3mm}
\end{table}
}

\input{tables/experiments}

\subsubsection{Baseline Methods}
We compare \nm with the following baseline methods on fault-indicating information extraction. 
\begin{itemize}[leftmargin=*, topsep=0pt]
\item TF-IDF~\cite{lee2008tfidf} is a keyword extraction method based on word frequency. We treat a log session as a corpus and each log as a document. The importance weights of words are calculated using TF-IDF, and the top-$k$ words with the highest weights are chosen as FID and FIP. We set $k=6$ as is close to the average length of FID and FIP.

\item TextRank~\cite{mihalcea2004textrank} is a summarization method based on textual similarity. We use Log2Vec~\cite{liu2019log2vec} to produce word embeddings, and then rank the words based on the importance scores computed by TextRank. We use the top-$k$ words with the highest score as FID and FIP where $k=6$.

\item LogSummary~\cite{meng2023logsummary} is a log summarization method, which first extracts summaries in triple format from log templates and then ranks the triples based on Log2Vec~\cite{liu2019log2vec}. We obtain the highest scoring one as the FID and FIP. 

\item ChatGPT~\cite{chatgpt} is a cutting-edge conversational AI based on large language models, which has been widely acknowledged for its groundbreaking abilities in text understanding. Specially, we consider a zero-shot and an In Context Learning (ICL) setting, where the former construct the prompt input with questions (\S~\ref{sec:method-stage-2-prompt}) and logs, and the latter additionally incorporates one more example as demonstrations. We use OpenAI APIs to obtain responses. Due to space constraints, we show the prompt in our repository~\cite{lofi}.
\end{itemize}

\vspace{-0.5mm}
\subsubsection{Metrics}
Log fault-indicating information extraction aims to automatically identify the correct FID and FIP strings in a format of word tokens from logs, which is a new task in log analysis. To evaluate the task, we utilize the \textit{F1 score (F1)}, which is widely used to measure the correctness of extracted answer strings with the reference~\cite{rajpurkar2016squad, glass2020span}. F1 score measures the average overlap between the predictions and ground truth FID or FIP. As the F1 score is computed the same as the Rouge-1~\cite{lin2004rouge}, a popular metric to evaluate textual summarization, we simply use F1 here. We treat each prediction and reference as bags of words and compute the \textit{Precision}, \textit{Recall}, and \textit{F1-score} of each example as follows: $precision=\frac{\#\;shared\;tokens}{\#\;prediction\;tokens}$, $recall=\frac{\#\;shared\;tokens}{\#\;ground\;truth\;tokens}$, $F1=\frac{2\times precision \times recall }{precision\;+\;recall}$. Finally, the scores are averaged over all examples for FID and FIP, respectively.

\vspace{-0.5mm}
\subsubsection{Implementation Details}
We conduct our experiments on a Linux GPU server with Intel Xeon 2.3GHz CPU and NVIDIA Tesla V100 16G GPU. We implement \nm with Python 3.9, PyTorch 2.0 and transformers 4.26.1. For the log selection module, we set 10 seconds as the default time to collect a session of anomalous logs. When training the pre-trained language models, we use AdamW \cite{adamw} optimizer with a learning rate of 5e-5 and linear scheduling with 5\% warm-up. The maximum input token length for UniXcoder is 512. We set the training batch size as 8 and train the model for 100 epochs. We merge the top-$3$ scoring spans to get predictions. In the online stage, we set the inference batch to 32. The time period for online anomaly detection is 10 seconds.

\vspace{-2.5mm}
\subsection{RQ1: Effectiveness of \nm}
\label{sec:experiment-rq1}
\noindent\textbf{Setup.} In this RQ, we evaluate the effectiveness of \nm in extracting fault-indicating information in an offline setting by comparing it with a range of baseline models. To simulate the process, we assume the anomalies are all correctly detected and construct the log sessions with a time period of 10 seconds. We fine-tune the models on the training set with randomly sampled 32 cases of \datasetA and \datasetB and evaluate on the remaining cases, respectively.

\noindent\textbf{Results.}
The evaluation results in terms of precision, recall, and F1 scores are shown in Table \ref{tab:result-overall}. 
From the results, we can find that: 
(1) \nm achieves high accuracy in recognizing fault-indicating information, with F1 scores of 87.4/80.6 for FID/FIP on \datasetA and 72.2/62.8 for FID/FIP on \datasetB, which show the effectiveness of our method. \nm's superior results stem from its design, which combines an efficient log selection algorithm, a robust pretrained language model, and a context-aware prompt-based tuning mechanism. Further comparisons of the three components can be found in \S~\ref{sec:experiment-rq2} and \S~\ref{sec:experiment-rq3}.
(2) \nm significantly outperforms all baseline methods in both \datasetA and \datasetB for FID and FIP, across all evaluation metrics. Specifically, in terms of F1, \nm surpasses the strongest baseline \textit{ChatGPT-ICL} by 81\% (an average of 75.8 over 41.9 across two datasets and two fault-indicating information). This demonstrates \nm's substantial superiority over existing baselines. Unsupervised methods, such as TF-IDF, TextRank, and LogSummary, which do not use pre-trained language models, struggle to understand the log semantics, leading to low F1 scores in recognizing fault-indicating information across both datasets. Among unsupervised methods, ChatGPT-Zeroshot stands out, which is probably because ChatGPT is a powerful large language models and can understand the log and question semantics. Although incorporating the in-context learning mechanism boosts ChatGPT-ICL's  performance, it still falls short of \nm by a large margin. The performance gap further emphasizes \nm's superiority.
(3) Comparing the results on \datasetA and \datasetB, we find the overall performance on \datasetA is better than that on \datasetB (84 versus 67.5 on average of F1 for \nm). This is probably because \datasetB is collected from an industrial application which has a more diverse set of faults and more complicated logs.  
(4) Comparing the results of FID prediction and FIP prediction on both datasets, we find that the overall accuracy of FID is higher than FIP (79.8 versus 71.7 on average of F1). This can be attributed to the difference in word distribution between FID and FIP, where the former are mainly written in natural language, and the latter often contain parametric words like HTTP requests and user IDs in hexadecimal format, which is more difficult to recognize. Despite \nm is based on UniXcoder, a language model pretrained with mixture of natural language and code, it still performs poorer on FIP than FID.

\vspace{-1mm}
\subsection{RQ2: Impacts of Log Selection}\label{sec:experiment-rq2}
\noindent\textbf{Setup.} In this study, we examine the impacts of log selection (LS) in \nm on the performance of fault-indicating information extraction. The LS module first selects severe logs based on their logging levels and then collects related logs based on embedding similarity. To this end, we completely remove LS module and substitute it with four different LS methods. Next, we fine-tune the PLMs using varied inputs on the same dataset split as RQ1, and compare \nm's performance against these LS variants. Besides F1 scores, we also report the selection accuracy (Acc.), considering an example correct if the chosen logs carry FID and FIP, and the compression ratio (CR), which is the percentage of log lines remaining after LS.

\begin{itemize}[leftmargin=*, topsep=0pt]
\item w/o LS: The log selection module is removed.
\item LS=Error: Logs with the ERROR logging level are selected.
\item LS=ErrorWarn: Logs with ERROR or WARN levels.
\item LS=Highest: Logs with the highest logging level. 
\item LS=HighestCtx: Logs with the highest level are initially selected. Contextual logs, which appear immediately before and after these logs, are then merged with the first step.
\end{itemize}

\input{tables/ablation_log_selection}

\noindent\textbf{Results.} The results in Table \ref{tab:result-ablation-ls} shows that:
(1) Our proposed log selection method in the full \nm model outperforms in five out of six metrics, highlighting its superiority. Specially, when comparing full \nm to \nm with LS=Highest and LS=HighestCtx, we find that mining semantic-similar logs can complement the hard level selection, allowing for more accurate fault-indicating information extraction by finding additional potential fault-related logs.
(2) Despite using logs at ERROR and WARN levels (LS=ErrorWarn) can cover 100\% FID and FIP in \datasetB, the performance still falls short as it struggle to filter out irrelevant logs, resulting in a high compression rate of 86.4\%. As compression rates vary greatly with logging level-based filtering due to varying software logging styles, an adaptive LS strategy combining level and semantic selection is a more effective solution.

\vspace{-1mm}
\subsection{RQ3: Impacts of Prompt-based Tuning}\label{sec:experiment-rq3}
\noindent\textbf{Setup.} \nm utilizes prompt-based tuning with PLMs to understand the semantic meaning of log contents and predict the span of FID and FIP. In this RQ, we evaluate the effectiveness of prompt-based tuning with different prompt designs and tuning methods. We also investigate the influence of PLMs by replacing UniXcoder to various PLMs. 
\begin{itemize}[leftmargin=*, topsep=0pt]

\item Prompt=LessInfo: A less informative hard prompt is used~\cite{gao2021making}, \ie \textit{``Fault-indicating descriptions in the following logs: "} for FID and \textit{``Fault-indicating parameters in the following logs: ''} for FIP.  
\item w/o Prompt: The prompt question are removed and only the logs are used to fine-tune a UniXcoder model. 
\item w/o Tuning: Directly apply UniXcoder without finetuning to predict the span of FID and FIP with the input of prompt questions and logs.
\item PLM=BERT: The PLM is replaced by BERT~\cite{devlin2019bert}, the first PLM for language understanding. 
\item PLM=RoBERTa: The PLM is replaced by RoBERTa~\cite{liu2019roberta}, an improved version of BERT with optimized hyperparameters and a larger training corpus.
\item PLM=CodeBERT: The PLM is replaced by CodeBERT~\cite{feng2020codebert}, the first PLM trained on a mix of natural language and code.

\end{itemize}

\noindent\textbf{Results.} The results shown in Table \ref{tab:result-ablation-tuning} reveal that:
(1) Among the prompt variants, prompt question performs the best, suggesting that it provides important context for this task, thereby enhancing PLMs' ability to pinpoint relevant fault-indicating information. Notably, the absence of prompts (w/o Prompt) leads to a substantial performance drop, especially for FIP, which sees a 51.6\%/37.0\% F1 drop on \datasetA/\datasetB compared to FID with only 28.8\%/7.8\% F1 drop. This difference in decline ratio can be attributed to the variances in word distribution between FID and FIP. Since FIP often contain technical terms and domain-specific vocabulary that are less common in the pre-training corpus, predicting them without the context provided by the prompt is more challenging. 
(2) \nm shows poor performance without fine-tuning, especially with FIP. This result shows the importance of fine-tuning, which can significantly improve \nm's performance, even when available training data is limited. 
(3) Among the PLMs, UniXcoder achieves the best performance. Owing to its training on a hybrid corpus of code and natural language with improved training objectives, UniXcoder can better understand the semantics of mixed text types. Consequently, it can more effectively interpret logs, which comprise natural language and code-like parameters, thus underscoring its superiority.

\input{tables/ablation_prompt_tuning}

\subsection{RQ4: How \nm Assists in Online Diagnosis?}\label{sec:experiment-rq4}
In this RQ, we conduct a user study to evaluate \nm's effectiveness and usefulness in an online setting, where streaming logs are continuously produced by the systems. 

\noindent\textbf{Setup.} To simulate the diagnosis process in production environments, we train \nm on all 88 examples in \datasetB and apply our online pipeline (\S~\ref{sec:method-online}) to new logs from 2023-03-03 to 2023-04-02 to extract fault-indicating information. The pipeline first identifies anomalous logs sessions using an anomaly detection algorithm, then employs \nm for extraction. Due to company policy, the total number of anomalies is not disclosed. 
For the user study, 50 fault examples are randomly sampled and evaluated by ten experienced engineers  from three different service teams within \cloud, averaging 3.6 years of experience. 

We begin our study by showing participants with full raw logs in the anomalous session, accompanied by extracted FID and FIP. They are then asked to rate the accuracy of extracted FID and FIP (Q1-Q2). After judging all examples, they are asked if automatic fault-indicating information extraction would help (Q3). Specially, Q1 and Q2 are rated on a 5-point Likert scale~\cite{mcleod2008likert}, with participants encouraged to provide explanations for their scores. We then summarize the results.

\noindent\textbf{Q1. Do FIDs accurately represent anomalous events?}  
The average rating of the FID's accuracy was 4.34, with a majority of 30 examples scoring 5, 12 examples scoring 4, 4 scoring 3, 3 scoring 2, and 1 scoring 1. 
Overall, we find that participants highly acknowledged the effectiveness of FID in summarizing logs and aiding diagnosis. Apart from false predictions, some examples with low scores were explained by the participants, such as \emph{containing redundant information} and \emph{erroneous splitting}.

\noindent\textbf{Q2. Do FIP accurately identify anomalous components?} 
The average rating was 4.02, with a majority of 34 examples scoring 5, 5 scoring 4, and 11 scoring 1. Overall, participants were positive about FIP's correctness, though one participant noted a recurring issue with unnecessary predicted parameters:

\begingroup
\addtolength\leftmargini{-2.4em}
\begin{quote}
\textit{``The description of "sql cannot be full or empty" is correct. But the example can have a void parameter since the root cause is not system-related, but user-related.''}
\end{quote}
\endgroup

\noindent\textbf{Q3. Would extracting fault-indicating information aid in fault diagnosis?} 
Notably, all participants agreed that automated extraction of fault-indicating information from logs would help. In addition to reducing time and efforts for diagnosis, some participants comment on other benefits, such as implying root causes and following mitigation steps: 

\begingroup
\addtolength\leftmargini{-2.4em}
\begin{quote}
\textit{``The extracted fault-indicating descriptions can represent root causes and are useful for troubleshooting.''}
\end{quote}
\endgroup

\begingroup
\addtolength\leftmargini{-2.4em}
\begin{quote}
    \textit{``The fault-indicating information can correlate to possible mitigation steps, \eg when I see the description of "service does not exist" and the parameter of "serviceId=...", I realize I can first try to restart the service to mitigate.''}
\end{quote}
\endgroup

\begingroup
\addtolength\leftmargini{-2.4em}
\begin{quote}
    \textit{``When is this tool scheduled to launch? I used to spend 3-4 hours mitigating a fault, but with this tool, I'll be able to save time on checking hundreds of logs to find run-time behaviour.''}
\end{quote}
\endgroup

Overall, engineers in \cloud acknowledge the value of identified fault-indicating information in assisting with diagnosis, reducing time and human efforts. These findings highlight the utility of automatic extraction of FID and FIP from logs, shedding light on future research to improve log analysis and fault diagnosis by mining more useful log information.

\section{Industrial Experience}
In this section, we share our experience of applying \nm to real-world cloud service systems in \cloud, aiming to show its usefulness. 
In \cloud, numerous services use logs to record system runtime behaviors during runtime, which are retrieved and analyzed when engineers diagnose a fault. However, when analysing logs, engineers face two main challenges. 
Firstly, due to the increasing scale and complexity of online service systems, a single fault can cause cascading failures which trigger sequential events in a short period of time, leading to multiple log sessions from various services~\cite{li2021fogofwar, yang2021aid}.
Secondly, to support diagnosis, software systems may generate extra log lines to record the specifics when problems arise~\cite{yuan2012logenhancer}, \eg software stack, resulting in tens to hundreds of logs per session. Consequently, despite the use of log anomaly detection to reduce logs for investigation, engineers still struggle with high log volumes. 
To address this, in \cloud, \nm has been integrated into the intelligent log analysis system that serves hundreds of microservices to improve reliability.
\nm processes anomalous log sessions and extracts fault-indicating information to highlight symptoms and problematic positions. 
This allows engineers to swiftly understand the fault and grasp its essence, eliminating the need to read hundreds of logs and allowing a greater focus on troubleshooting and root cause analysis. 
In the following, we present two primary usage scenarios of identified fault-indicating information in \cloud: \textit{rapid diagnosis} and \textit{alert configuration}.

\noindent\textbf{Rapid Diagnosis.}
Identifying fault-indicating information enables online service providers to diagnose and recover from incidents rapidly. Figure \ref{fig: workflow} shows the diagnosis workflow. 
Upon realizing a failure, engineers first determine relevant services and gather anomalous logs from these services. Then they manually review these logs to pinpoint incident-related information, \eg failure events and faulty devices, determine the root cause, and implement a mitigation plan. However, this process can be time-consuming and potentially inaccurate, especially in large-scale cloud systems with numerous simultaneously events. 
\nm automates this by extracting and highlighting FID and FIP, providing an immediate snapshot of anomalous actions and variables, thus significantly reducing the time to identify the root cause.

\noindent\textbf{Alert Configuration.}
Another use case of \nm is configuring alerts with summarized fault-indicating information from logs during system monitoring. 
Alerting is commonly used in software monitoring to indicate potential issues that need attention. The current monitoring raise alerts based on predefined conditions or thresholds, such as specific log occurrences or detected anomalies~\cite{li2021fogofwar,li2022going, yang2022characterizing, kuang2024knowledge}. However, the alert names are typically set by predefined templates, offering limited actionable guidance to engineers.
\nm streamlines this process by automatically creating alert contents with extracted fault-indicating information, which can specify the events that occurred, and the associated objects or variables. The summary provides a clear overview of the anomaly and actionable alerts to the SRE team, improving the overall efficiency of the entire incident response process.

%% file: tables/experiments.tex
\begin{table*}[h]
  \centering
    \caption{Experimental results (\%) of log fault-indicating information extraction}
\vspace{-3mm}
    \resizebox{0.8\textwidth}{!}{
    \begin{threeparttable}
    \begin{tabular}{lccc|ccc|ccc|ccc}
    \toprule
    \multicolumn{1}{c}{\multirow{2}[1]{*}{Method}} & \multicolumn{3}{c}{\datasetA-FID} & \multicolumn{3}{c}{\datasetA-FIP} & \multicolumn{3}{c}{\datasetB-FID} & \multicolumn{3}{c}{\datasetB-FIP} \\
          & \multicolumn{1}{l}{Precision} & \multicolumn{1}{l}{Recall} & \multicolumn{1}{l}{F1} & \multicolumn{1}{l}{Precision} & \multicolumn{1}{l}{Recall} & \multicolumn{1}{l}{F1} & \multicolumn{1}{l}{Precision} & \multicolumn{1}{l}{Recall} & \multicolumn{1}{l}{F1} & \multicolumn{1}{l}{Precision} & \multicolumn{1}{l}{Recall} & \multicolumn{1}{l}{F1} \\
    \midrule
    TF-IDF$^*$ &  3.4    &   2.6    &   2.8  &  2.8   &  1.4     &  1.8    & 5.3  & 4.2  &  4.4   & 2.5 & 2.4   &  2.4   \\ 
    TextRank$^*$ &    12.8   &   12.8    &   12.3   &  0.0    &   0.0    &   0.0 &  17.9 & 18.0  &  17.1    &   5.2 & 3.6  &  4.2 \\
    LogSummary$^*$ &    4.5  &  5.4  &   4.5   &   3.6 &  1.1 &    1.7  &     16.2 & 13.5  &  14.2 &  6.9   &  4.9  &  5.5    \\
    ChatGPT-Zeroshot$^*$ &  \underline{59.6} & 30.1  &  38.2 &    9.7  & 1.3  &   2.2  &   \underline{47.2} &	29.9 & 33.2 &  32.1 &	33.3	& 32.2  \\
    ChatGPT-ICL &     53.3  & \underline{51.6}  &  \underline{49.6}   &  \underline{46.5} & \underline{44.4}  &  \underline{44.9} &  45.1 &	\underline{33.3} & \underline{35.9} &  \underline{41.3} & \underline{38.3} & \underline{37.0}   \\
    \midrule
    \textbf{\nm} (ours) & \textbf{87.4} & \textbf{87.6} & \textbf{87.4} & \textbf{80.6} & \textbf{80.6} & \textbf{80.6} & \textbf{73.8} & \textbf{72.0} & \textbf{72.2} & \textbf{70.0}    & \textbf{60.9}& \textbf{62.8} \\
    \bottomrule
    \end{tabular}%
    \begin{tablenotes}
        \small
        \item[1] We use $*$ to denote unsupervised methods, others are supervised ones.
    \end{tablenotes}

    \end{threeparttable}
}
\vspace{-4mm}
  \label{tab:result-overall}%
\end{table*}

%% file: tables/ablation_log_selection.tex
\begin{table}[t]
  \centering
  \caption{Results of different log selection (LS) methods  (\%) }
\vspace{-3mm}
    \resizebox{0.485\textwidth}{!}{
\begin{tabular}{lrrrr|rrrr}
\toprule
    & \multicolumn{4}{c}{\datasetA} & \multicolumn{4}{c}{\datasetB} \\
    \multicolumn{1}{r}{} & \multicolumn{1}{l}{CR} & \multicolumn{1}{l}{Acc.} & \multicolumn{1}{l}{F1-FID} & \multicolumn{1}{l}{F1-FIP} & \multicolumn{1}{l}{CR}  & \multicolumn{1}{l}{Acc.} &  \multicolumn{1}{l}{F1-FID} & \multicolumn{1}{l}{F1-FIP} \\
\midrule
    \multicolumn{1}{l|}{\textbf{Full \nm}} & 39.9 & \textbf{100.0} & \textbf{87.4}  & \textbf{80.6} & 62.7 & 98.2  & \textbf{72.2}  & \textbf{62.8}  \\
    \multicolumn{1}{l|}{- LS=HighestCtx} & 38.1 & 87.2  & 77.6  & 78.2 & 59.1 & 91.1  & 67.6  & 58.9  \\
    \multicolumn{1}{l|}{- LS=Highest} & 34.5 & 76.9  & 69.9  & 72.2 & 48.2 & 91.1  & 67.0  & 60.8  \\
    \multicolumn{1}{l|}{- LS=ErrorWarn} & 7.9& 71.8  & 65.4  & 59.3 & 86.4 & \textbf{100.0} & 61.0  & 29.5  \\
    \multicolumn{1}{l|}{- LS=Error} &  4.0  & 23.1  & 16.7  & 13.9 & 44.6 & 85.7  & 64.0  & 56.5  \\
    \multicolumn{1}{l|}{- w/o LS} & \multicolumn{1}{r}{-} & \multicolumn{1}{r}{-} & 76.9  & 75.6 & \multicolumn{1}{r}{-} & \multicolumn{1}{r}{-} & 50.2  & 43.7  \\
\bottomrule
\end{tabular}%
}
\vspace{-5mm}
  \label{tab:result-ablation-ls}%
\end{table}%

%% file: tables/ablation_prompt_tuning.tex
\begin{table}[t]
  \centering
  \caption{Results of prompt-based tuning variants (\%)}
\vspace{-3mm}
    \resizebox{0.34\textwidth}{!}{
    \begin{tabular}{lrrrr}
    \toprule
    & \multicolumn{2}{c}{\datasetA} & \multicolumn{2}{c}{\datasetB}  \\
    & \multicolumn{1}{l}{F1-FID} & \multicolumn{1}{l}{F1-FIP} & \multicolumn{1}{l}{F1-FID} & \multicolumn{1}{l}{F1-FIP} \\
\midrule
    \textbf{Full \nm}  & \textbf{87.4} & \textbf{80.6} & \textbf{72.2} & \textbf{62.8} \\
    - Prompt=LessInfo   & 81.3 & 78.3 & 68.6 & 57.9 \\
    - w/o Prompt & 58.6 & 29.0 & 64.4 & 25.8 \\
    - w/o Tuning & 34.8 & 1.8  & 11.3 & 5.6 \\
\midrule
    - PLM=CodeBERT & 81.5 & 76.4 & 69.6 & 49.5 \\
    - PLM=RoBERTa & 83.8 & 78.3 & 64.7  & 41.4 \\
    - PLM=BERT & 77.5 & 78.7  & 65.4 & 14.6 \\

\bottomrule
\end{tabular}%
}
\vspace{-5mm}
  \label{tab:result-ablation-tuning}
\end{table}

%% file: content/06_discussion.tex
\vspace{-0.7em}
\section{Threats to Validity}
\noindent\textbf{Internal Threats} \nm is designed to extract fault-indicating information from logs when anomalies occur. Consequently, it may struggle with normal log sessions due to different data distributions and absence of fault-indicating information, \eg error status codes, error messages, faulty devices, \etc. Nevertheless, since normal logs are less critical and do not require investigation unless anomalies are present, we consider the model's poor performance with normal logs acceptable.

\noindent\textbf{External Threats} In this paper, the task and method are inspired by real-world fault mitigation practices with logs in \cloud. One may concern \nm's applicability to other systems from different companies. Since logs have long been recognized as critical for software maintenance, and \cloud is a large-scale cloud company providing hundreds of online services for millions of users, we believe \cloud's practices are representative.
Due to privacy and security concerns, we cannot access data from other companies, so \nm is only evaluated on \cloud logs. However, the proposed approach can be trained with a small number of labeled data, making it feasible to apply \nm to anomalous logs from other systems.

Another threat arises from the labeled data. \nm relies on a small set of labeled log fault-indicating information for training and evaluation. The annotation requires engineers to manually inspect the log sessions and history mitigation records and mark the information span from the logs. Limited by engineers' varying experience, the label may not be entirely accurate. However, the engineers involved are in charge of cloud monitoring in \cloud and have rich experience in fault localization and mitigation using logs. They can also search online and discuss with colleagues to reach a consensus. Thus, we believe the amount of inaccurate labels is small (if any). By introducing extra annotated data to cover more anomalous scenarios, our method is expected to predict fault-indicating information more precisely.

%% file: content/07_related_work.tex
\vspace{-2mm}
\section{Related Works}
\vspace{-0.5mm}
\subsection{Logs Analysis for Failure Diagnosis} 
Logs are essential for diagnosing online service systems. To reduce diagnosis efforts, various tools have been proposed to mine informative content from large volumes of logs at different granularities~\cite{he2021survey,he2022empirical}.

One research direction focuses on mining a small set of log messages for diagnosis.  
Anomaly detection \cite{le2022loghowfar} is a key task that identifies a session of logs deviating from normal behaviors. The sessions are usually split by a  time period or fixed length and then judged normal or not based on various techniques, including traditional machine learning~\cite{xu2009largescale, liang2007failure, lin2016logclustering, zhao2021empiricalLogAD, yang2021semi, zhang2023semi}, deep learning~\cite{du2017deeplog, nedelkoski2020selfattention, zhang2019robustAD, zhang2022deeptralog}, 
and language models~\cite{le2021withoutparse, almodovar2022can}. As a session can contain numerous continuous logs and noise, some works propose to highlight lines that are most likely to reveal problems by clustering logs. LogFaultFlagger~\cite{amar2019LogFaultFlagger} flags problematic log lines for human inspection, while 
LogSed~\cite{jia2017logsed} uses a time-weighted control flow graph to identify problematic logs across multiple software threads. Onion~\cite{zhang2021onion} identifies three aspects of incident-indicating logs (\ie consistency, impact, and bilateral difference) and locate these logs with clustering. However, due to the complexity and verbosity of log content, manually examining lines of logs remains time-consuming for engineers to understand runtime status and diagnose system problems~\cite{dogga2019debuggingAssistant, sui2023logkg}.

Another line of research focuses on extracting useful information from log messages at a finer granularity. Log parsing \cite{khan2022guidelines,zhang2023parsingSurvey} identifies variables in logs by converting logs into static templates and dynamic variables~\cite{makanju2009IPLoM, he2017drain,zhang2017FTtree,liu2022uniparser,wang2022spine,le2023logppt,tao2022logstamp, dai2020logram}. 
But these variables can not be directly used for diagnosis as the corresponding categories and importance levels are unaware \cite{li2023valb}. To address this, SemParser \cite{huo2021semparser} and VALB \cite{li2023valb} explicitly identify variables and corresponding concept types (\eg concepts like "instance" and "server") during parsing. 
However, these two works are used for log parsing, and the identified variable and types may not relate to failures. 
LogSummary \cite{meng2023logsummary} is a closely related work to ours. They extract multiple triples from logs, \ie ("entity", "event", "relation"), to represent a log sequence and improve readability. However, these simple triples may not capture the complex semantics of logs, and users are still unaware of severe events and instances to help diagnosis.

\vspace{-2.5mm}
\subsection{Language Models for Log Analysis}
\vspace{-0.5mm}
Pre-trained language models (PLM) (\eg BERT~\cite{devlin2019bert}) have significantly advanced many software engineering tasks~\cite{lu1codexglue} due to the strong ability to learn the semantic information of the textual input. Based on the idea that logs are language sequences \cite{zhang2019robustAD}, many studies applied PLMs to log analysis. Swisslog \cite{li2020swisslog} tunes BERT to encode log messages for anomaly detection. 
Ott et al. \cite{ott2021robust} examine the effectiveness of various PLMs for anomaly detection. NeuralLog~\cite{le2021NeuralLog} directly feeds raw logs to PLMs to avoid the issues of out-of-vocabulary words and semantic misunderstandings. Apart from anomaly detection, PLMs are also used for log parsing. LogStamp~\cite{tao2022logstamp} treats log parsing as a sequence labeling problem, fine-tuning BERT to judge whether tokens are parameters. LogPPT~\cite{le2023logppt} explores prompt-based tuning on RoBERTa~\cite{liu2019roberta} to identify templates and parameters. 

However, these studies fail to provide actionable insights to assist on-site engineers in diagnosing faults. Our work, \nm trains a PLM to learn semantics from a few labeled examples with prompt-based tuning and can effectively identify fault-indicating descriptions and parameters from logs for diagnosis.

%% file: content/08_conclusion.tex
\section{Conclusion}

In conclusion, this paper has addressed the challenge of extracting fault-indicating information from anomalous log sessions to aid fault diagnosis and reduce mitigation efforts in large-scale software systems. We conducted a preliminary study on log-based troubleshooting practices at \cloud, and identified two types of information engineers typically prioritize, \ie fault-indicating descriptions (FID) and fault-indicating parameters (FIP). Motivated by this finding, we proposed a two-stage approach, \nm, which performs coarse-grained filtering in the first stage and leverages a pre-trained language model with a novel prompt-based tuning method in the second stage to extract fine-grained fault-indicating information. Our evaluation of \nm on \datasetA and industrial datasets demonstrated significant improvements over baseline methods. Additionally, our user study and successful deployment to \cloud provided further evidence for the usefulness of our proposed method.

\section{Acknowledgement}\label{sec:Acknowledgement}

The work described in this paper was supported by the Research Grants Council of the Hong Kong Special Administrative Region, China (No. CUHK 14206921 of the General Research Fund) and Fundamental Research Funds for the Central Universities, Sun Yat-sen University (No. 76250-31610005).